\documentclass[draft,a4paper]{article}
\usepackage[
    pdfauthor={Erica Bertolini, Hyungrok Kim},
    pdftitle={Strings as Hyper-Fractons},
    pdfsubject={hep-th,cond-mat},
    breaklinks=true,
    final
]{hyperref}
\usepackage{amsmath,amssymb,amsthm,mathtools,orcidlink,booktabs,cleveref,rotating,pdflscape}
\usepackage[russian,greek,british]{babel}
\usepackage[varbb]{newpxmath}
\usepackage[osf]{newpxtext}
\usepackage{CJKutf8}
\usepackage{float}
\usepackage{multicol}
\usepackage{multirow}
\newtheorem{proposition}{Proposition}
\title{Strings as Hyper-Fractons}
\author{Erica Bertolini\orcidlink{0000-0003-0773-7526}\footnote{School of Theoretical Physics, Dublin Institute for Advanced Studies, 10 Burlington Road, \textsc{d04~c932}, Dublin, Ireland.\\Scoil na Fisice Teoiriciúla, Institiúid Ard-Léinn Bhaile Átha Cliath, 10 Bóthar Burlington, \textsc{d04~c932}, Baile Átha Cliath, Éire.}\\\href{mailto:ebertolini@stp.dias.ie}{\texttt{ebertolini@stp.dias.ie}}\\Hyungrok Kim (\begin{CJK*}{UTF8}{bsmi}金炯錄\end{CJK*})\orcidlink{0000-0001-7909-4510}\footnote{Department of Mathematics and Theoretical Physics, University of Hertfordshire, Hatfield, Hertfordshire \textsc{al10~9ab}, United Kingdom.}\\\href{mailto:h.kim2@herts.ac.uk}{\texttt{h.kim2@herts.ac.uk}}}
\renewcommand\delta\deltaup
\renewcommand\Sigma\Sigmait
\begin{document}
\maketitle
\begin{abstract}
We systematically examine all possible Gauss laws obeying spatial rotation symmetry, characterising the corresponding conserved charges. In the case of conserved higher moments, this gives rise to fractonic behaviour. We show that many Gauss laws, including those arising from \(p\)-form electrodynamics, in fact, produce an infinite tower of conserved moments, which we dub \emph{hyper-fractonic}. In hyper-fractonic systems, a finite number of charged particles cannot be mobile due to an inability of fulfilling the infinite number of conservation laws with a finite number of degrees of freedom. Instead, mobile charged objects must have an infinite number of degrees of freedom. In particular, the strings and branes naturally coupling to \(p\)-form potentials provide an example of hyper-fractonic matter.
\end{abstract}
\tableofcontents
\section{Introduction and summary}
Fractons \cite{Vijay:2015mka,Vijay:2016phm,Pretko:2020cko,Nandkishore:2018sel,You:2024zyk} are a fascinating phenomenon, originally discovered in lattice spin glasses \cite{Chamon:2004lew} and quantum information \cite{Haah:2011drr,Bravyi:2013ort} but subsequently also found in gauge theories \cite{Pretko:2016kxt,Pretko:2016lgv}, Carrollian theories \cite{Figueroa-OFarrill:2023vbj,Figueroa-OFarrill:2023qty}, elastic media \cite{Pretko:2017kvd,Pretko:2019omh,Gromov:2020yoc,Yuan:2019geh,Wang:2019mtt,Grosvenor:2021rrt,Glodkowski:2022xje,Tsaloukidis:2023jmr} and many more systems \cite{Yan:2018nco,You:2019bvu,Seiberg:2020bhn,Seiberg:2020wsg,Slagle:2020ugk,Caddeo:2022ibe,Huang:2023zhp,Perez:2023uwt}

Indeed, as stated in \cite{Nandkishore:2018sel} ``the fracton frontiers sit at the confluence of multiple streams of research in theoretical physics'', and thus, depending on the particular area of physics,  there are variations about the technical definition of a fractonic behaviour. The very definition of the word ``fracton'' dates back to \cite{Vijay:2015mka}, in which ``the fundamental excitations are termed fractons, as they behave as fractions of a mobile quasiparticle'', immediately followed in \cite{Vijay:2016phm} by ``\dots\ a generalised Gauss law that characterises the fracton topological phase''.

What is conserved is thus not only the total charge but also additional higher moments of the charge \cite{Gromov:2018nbv}. As a result, single charged particles must be immobile \cite{Pretko:2016kxt,Pretko:2016lgv,Nandkishore:2018sel,Pretko:2020cko}: a single particle moving around may have constant charge, but the dipole moment will change. If the charge and the dipole moment are conserved, then dipoles (pairs of particles of opposite charges) may be mobile. This has been for years the common thread for defining fracton theories: limited mobility and electromagnetic-like behaviour (generalized Gauss and multipole conservation). The literature is divided into two main classes \cite{Pretko:2020cko,Nandkishore:2018sel}: lattice models (e.g.~\cite{Chamon:2004lew,Haah:2011drr,Vijay:2015mka,Vijay:2016phm}) and tensor gauge theories (e.g.~\cite{Pretko:2016kxt,Pretko:2016lgv}), and in this paper we are concerned with the latter class. It is interesting to observe that these two classes describe inherently different fracton models. For instance the first case defines four-dimensional gapped theories, with possible fractal structure for the case of the so-called Type~II models \cite{Haah:2011drr,Pretko:2020cko}. On the other hand tensor gauge theories are gapless, and it has been possible to build \((2+1)\)-dimensional models as well \cite{Nandkishore:2018sel}, which are physically relevant for their connection with elasticity theory \cite{Pretko:2017kvd}. The definition of fracton theories then stretched even further when fractonic mobility features has been identified also in gravity \cite{Pretko:2017fbf} (thrugh concervation of centre of mass), and Carrollian theories \cite{Figueroa-OFarrill:2023vbj} (through the lightcone limit $c\to0$). The mobility constraints are typically imposed through generalised Gauss laws. Different kinds of Gauss laws give different kinds of conservation laws, which imply different kinds of mobility of particles, thus creating fractons, lineons, planons, etc.
The higher the conserved moments, the more particles are needed to have mobility. In this context our paper follows the paradigm of \cite{Pretko:2016kxt,Pretko:2016lgv} and tensor gauge theories of fractons in general, where Gauss law constraints provide the key ingredient for limited mobility through dipole-moment (or more generally multipole-moment) conservation.

In fractonic gauge theories, the prototypical \((3+1)\)-dimensional examples \cite{Pretko:2016lgv} are defined by a symmetric tensor of rank \(2\) whose conjugate momentum (an electric-like field $E^{ij}$) is required to satisfy some particular Gauss law. This Gauss law is at the core of the limited mobility of the fracton quasiparticle through dipole (or quadrupole) moment conservation. Thus in the paradigmatic example of the scalar charge theory of fractons \cite{Pretko:2016lgv}, one has the Gauss law
	\begin{equation}
	\partial_i\partial_jE^{ij}=\rho
	\end{equation}
	for the electric field \(E^{ij}\) and charge density \(\rho\),whose consequence is to constrain isolated charges to be immobile while dipole bound states $d^i\sim x^i\rho$ are free to move \cite{Pretko:2016lgv,Pretko:2020cko,Nandkishore:2018sel}. Different constraints can change the nature of the conserved charges and their mobility. For instance, in the vector charge theory \cite{Pretko:2016lgv}, for which the Gauss constraint is instead
	\begin{equation}
	\partial_jE^{ij}=\rho^i,
	\end{equation}
the (now vector-valued) charge is a lineon (i.e.\ it can move only along a line). Imposing a tracelessness condition on the electric field, i.e.\ $E^i{}_i=0$, further limits the mobility of the above mentioned quasiparticles and their bound states. The fractonic phenomenology of these \((3+1)\)-dimensional tensor gauge theories is summarised in the table below.
\begin{table}[H]
	\resizebox{1\columnwidth}{!}{%
		\bgroup
		\setlength\tabcolsep{15pt}
		\def\arraystretch{1.5}{
			\begin{tabular}{cccc}\toprule
				Model&Gauss law&Conserved quantities&Particle content\\\midrule
				\multirow{2}{*}{Scalar charge theory} & \multirow{2}{*}{$\partial_i\partial_jE^{ij}=\rho$} &$\rho$ charge&monopole: fracton;\\[-5px] 
				&&$\vec x\rho$ dipole &dipole: free\\[1em]
				\multirow{2}{*}{Traceless scalar charge theory} & {$\partial_i\partial_jE^{ij}=\rho$} &$\rho\ ;\ \vec x\rho$
&monopole: fracton;\\[-5px] 
				&$E^i_{\ i}=0$&$x^2\rho$ $\sim$ quadrupole &dipole: planon\\[1em]
				\multirow{2}{*}{Vector charge theory} & \multirow{2}{*}{$\partial_iE^{ij}=\rho^j$} &$\vec\rho$ charge ($\sim$dipole)&\multirow{2}{*}{monopole: lineon}\\[-5px] 
				&&$\vec x\wedge\vec\rho$ angular momentum &\\[1em]
				\multirow{2}{*}{Traceless vector charge theory} & {$\partial_iE^{ij}=\rho^j$} &$\vec\rho\ ;\ \vec x\wedge\vec\rho$&\multirow{2}{*}{monopole:  fracton}\\[-5px] 
				&$E^i_{\ i}=0$&$\vec x\cdot\vec\rho$ ($\sim$ quadrupole)$\ ;\ (\vec x\cdot\vec\rho)\vec x+\frac 1 2 x^2\vec\rho$ &\\
				\bottomrule
			\end{tabular}
		\egroup}
	}
	\caption[Fracton phenomenology]{\footnotesize{\((3+1)\)-dimensional fracton phenomenology of rank-2 tensor gauge theories \cite{Pretko:2016lgv}. In the above, ordinary fractons, lineons, and planons are particles constrained to a point, line, or plane in space, respectively.}}\label{fracton_pheno}
\end{table}
The study of more complex cases is already complicated in \(3+1\) dimensions and  even more so in other dimensions. It is the aim of this paper to investigate, through representation-theoretic tools, all possible (Gauss-like) constraints in a general number of spatial dimensions and for gauge fields of arbitrary rank, and classify them through the associated conserved quantities and quasiparticle behaviour. The analysis allows us to uncover new kinds of possible fractonic behaviour. For instance, what would happen if there were infinitely many conservation laws? One might be tempted to guess that such systems must be unphysical as no finite system of charged particles can withstand such a straitjacket. However, in this paper we show that the familiar theory of \(p\)-form electrodynamics (for \(p>1\)) --- and, more generally, higher gauge theory \cite{Borsten:2024gox} --- provide examples of such \emph{hyper-fractonic} systems: there are infinitely many charges, requiring infinitely many particles which must dance around in a coordinated fashion so as to scrupulously observe all conservation laws and, in the process, arrange themselves into a \((p-1)\)-brane as is familiar from string theory \cite{Polchinski:1998rq,Polchinski:1998rr,Becker:2006dvp}.
Such \(p\)-forms and higher gauge theories have been discussed in a fractonic context before, for instance in \cite{Shenoy:2019wng}, but this is the first time that infinite towers of conserved moments have been discussed in higher gauge theory to our knowledge. Interestingly, a particular case of the hyper-fractonic behaviour of the Laplace equation was observed in \cite{Hart:2021gre} in the context of hydrodynamics, providing another physical example of our analysis. Note that the presence of this infinite tower of conserved moments may, but need not, imply integrability --- for example, in geometric hydrodynamics, the even-dimensional incompressible Euler equation admits an infinite number of conserved enstrophy-type integrals despite not being integrable \cite{arnold}; and in many Poincaré-invariant systems, the Coleman--Mandula theorem requires higher conserved moments to vanish \cite{Coleman:1967ad,Mandula:2015}.
Also, note that lattice regularisation and finite-size effects may break part of the tower of conservation laws \cite{Hart:2021gre,Khudorozhkov:2024gzo}.

In addition, we provide a discussion of general Gauss laws in an arbitrary number of spacetime dimensions, assuming only spatial rotation and spacetime translation symmetry. A scan of all possible Gauss laws in low rank shows that such a hyper-fractonic behaviour is surprisingly common (see \cref{ssec:low_rank_scan}), including the Poisson equation (\cref{ssec:polyharmonic}), where the infinite tower of conserved moments corresponds to harmonic polynomials). Even in the non-hyper-fractonic case, we provide a systematic analysis and classification of the possible conserved moments using representation-theoretic tools.

In some cases action principles have been proposed, such as the covariant fracton model of \cite{Blasi:2022mbl,Bertolini:2022ijb,Bertolini:2023juh,Bertolini:2023sqa,Bertolini:2024gzx} or the \((n,k)\)-Maxwell models of \cite{Shenoy:2019wng}, and for \(p\)-form electrodynamics, but in general the obtainability of a given Gauss law from an action principle is not straightforward, as it is strictly model-dependent.
For instance, a symmetry should be imposed, under which the electric field must be invariant, being a physical observable. Moreover it also depends on the field content in terms of which the electric field is typically the conjugate momentum. In the context of this paper we will only discuss all possible Gauss laws, which could then be adapted to the models of interest.

Similar to the usual fractonic case \cite{Griffin:2014bta,Gromov:2018nbv}, a hyper-fractonic theory admits a multipole algebra. The approach presented in this paper starts from the analysis of all the possible Gauss laws in any dimensions and with gauge field of arbitrary rank, which then leads to the classification of conserved moments and fractonic behaviour. In contrast, Ref. \cite{Gromov:2018nbv} has the opposite approach, $i.e.$ from the multipole algebra Gauss laws are identified. What is relevant to remark is that, as hoped, the results intersect, and enrich one another, in our case for example through the interpretation of the new hyperfractonic behaviour in the classification. Finally we observe that for Poincaré-invariant theories meeting the assumptions of the Coleman--Mandula theorem  \cite{Coleman:1967ad,Mandula:2015} (such as \(p\)-form electrodynamics), the Coleman--Mandula theorem forces the vanishing of all conserved moments whose generators do not commute with the Poincaré group.\\

The paper is organised as follows. In Section \ref{sec:general_conserved_moments}, we classify all possible Gauss laws and consequent constraints assuming spatial rotational symmetry using representation theory, from which three possible behaviours are observed: non-fractonic, fractonic, and hyper-fractonic. In Section \ref{sec:fract}, we apply the results to examples of fractonic theories known in the literature, such as the traceless scalar and vector charge theories of Table~\ref{fracton_pheno}. In Section \ref{sec:hyper_fracton}, we consider examples of the hyper-fractonic case, most of which are either new or whose fractonic features have not been previously appreciated in the literature. Section~\ref{sec:final_remarks} presents a summary and outlook of our results.

\subsection{Notation and conventions}
We use \((\dotso)\), \([\dotso]\), and \(\langle\dotso\rangle\) for normalised total symmetrisation, total antisymmetrisation, and total traceless symmetrisation. For example, on rank 2 and 3 tensors,
\begin{subequations}
\begin{align}
    X^{(ij)} &= \frac12(X^{ij}+X^{ji}) \\
    X^{[ij]} &= X^{ij}-X^{ji} \\
    X^{\langle ij\rangle} &= X^{(ij)} - \frac1dX^{kl}\delta_{kl}\delta^{ij}\\
    X^{(ijk)} &= \frac1{3!}(X^{ijk}+X^{ikj}+X^{jki}+X^{jik}+X^{kij}+X^{kji}) \\
    X^{[ijk]} &= \frac1{3!}(X^{ijk}-X^{ikj}+X^{jki}-X^{jik}+X^{kij}-X^{kji}) \\
    X^{\langle ijk\rangle} &= X^{(ijk)} - \frac{\delta_{lm}}{d+2}\left(\delta^{ij}X^{(klm)}+\delta^{jk}X^{(ilm)}+\delta^{ki}X^{(jlm)}\right).
\end{align}
\end{subequations}
We will consider finite-dimensional irreducible representations of the simple Lie group \(\operatorname{SO}(d)\), which can be classified using the theory of highest weights. Given representations \(R_1\) and \(R_2\), we denote their tensor product as \(R_1\otimes R_2\), direct sum as \(R_1\oplus R_2\), and (if \(R_1\) and \(R_2\) are irreducible) the highest-weight subrepresentation of \(R_1\otimes R_2\) as \(R_1\vee R_2\). The \(n\)th symmetric tensor power of a representation is denoted \((-)^{\odot n}\), and the \(n\)th antisymmetric tensor power of a representation is denoted \((-)^{\wedge n}\), and the highest-weight subrepresentation of the symmetric power is denoted \((-)^{\vee n}\). The defining \(d\)-dimensional representation of \(\operatorname{SO}(d)\) is denoted as \(\square\) (similar to a single-cell Young tableau); thus, \(\square^{\wedge k}\) is the totally antisymmetric \(k\)-tensor, and \(\square^{\odot k}\) is the totally symmetric \(k\)-tensor, and \(\square^{\vee k}\) is the totally traceless totally symmetric \(k\)-tensor.

We use superscript or subscript \(i,j,a,A\) for labelling components, except that \(\Delta^m\) and \(|x|^{2n}\) denote exponentiation. When we discuss Lorentzian signature, our metric is mostly plus, and we use Greek indices \(\mu,\nu,\dotsc\). 

\section{Conserved charges for general representations}\label{sec:general_conserved_moments}
We first discuss a general Gauss law and the conserved multipole moments it implies, using representation theory of the rotation group \(\operatorname{SO}(d)\). We work in \(d\) flat spatial dimensions.

Suppose that we have an electric-like field (which we henceforth simply call `electric field', regardless of the physical interpretation) \(E^A\) that transforms as a representation \(R\) of \(\operatorname{SO}(d)\), where the indices \(A,B,\dotsc\in\{1,\dotsc,\dim R\}\) label the components of \(R\). Suppose that the charge density \(\rho^a\) transforms as a representation \(S\) of \(\operatorname{SO}(d)\), where the indices \(a,b,\dotsc\in\{1,\dotsc,\dim S\}\) label the components of \(S\).
By a \emph{Gauss law} we mean a general relation between \(E^A\) and \(\rho^a\) involving differential operators of the form
\begin{equation}\label{eq:gauss_law_ansatz}
    Y_A^{i_1\dotso i_k a}\partial_{i_1}\dotsm\partial_{i_k}E^A=\rho^a,
\end{equation}
where \(S\) is a subrepresentation
\begin{equation}
    S \subset \square^{\odot k} \otimes R,
\end{equation}
in which  \(\square^{\odot k}\) is the totally symmetric tensor representation of rank \(k\), which appears as the \(\operatorname{SO}(d)\)-representation of \(\partial_{i_1}\dotsm\partial_{i_k}\), 
and \(Y\) is the projection operator
\begin{equation}\label{eq:Y-definition}
    Y\colon \square^{\odot k}\otimes R \to S.
\end{equation}

A charge moment in some spatial region \(U\subset\mathbb R^d\) is, in general, a quantity of the form
\begin{equation}\label{eq:moment_general_form}
    P^\alpha_{j_1\dotso j_la}\int_U\mathrm d^dx\,x^{j_1}\dotsm x^{j_l}\rho^a,
\end{equation}
where \(P\) is a projector to a subrepresentation \(T\subset\square^{\odot l}\otimes S\) and \(\alpha\) is an index for \(T\).
Some of the above moments \eqref{eq:moment_general_form} obey \emph{conservation laws} if through \eqref{eq:gauss_law_ansatz} they can be expressed as local surface integrals involving the electric field \(E\):
\begin{equation}
    Q = \oint_{\partial U}\mathrm d^{d-1} \Phi(E),
\end{equation}
where \(\Phi\) is some local function of \(E\), which we interpret as the flux through the boundary \(\partial U\).
Thus, any change of \(Q\) inside a bounded spatial region \(U\subset\mathbb R^d\) corresponds to a flux \(\Phi(E)\) through its boundary \(\partial U\), leading to a local conservation law \cite{Pretko:2016kxt,Pretko:2020cko}.

By using the general Gauss law \eqref{eq:gauss_law_ansatz} in the definition of the charge moment \eqref{eq:moment_general_form}, we may explicitly express the total charge moment on some spatial region \(U\subset\mathbb R^d\) as
\begin{align}
    Q^\alpha&\coloneqq\int_U\mathrm d^dx\,P^\alpha_{j_1\dotso j_la}x^{j_1}\dotsm x^{j_l}\rho^a\notag\\
    &=P^\alpha_{j_1\dotso j_la}Y^{i_1\dotso i_ka}_A\int_U\mathrm d^dx\,x^{j_1}\dotsm x^{j_l}\partial_{i_1}\dotsm\partial_{i_k}E^A\label{eq:conservation_derivation}\\
    &=\begin{cases}(-1)^k\binom lkP^\alpha_{j_1\dotso j_la}Y^{i_1\dotso i_ka}_A\int_U\mathrm d^dx\,\delta_{(i_1}^{(j_1}\dotsm\delta^{j_k}_{i_k)}x^{j_{k+1}}\dotsm x^{j_l)}E^A&\text{if \(l\ge k\)}\\
    0 & \text{if \(l< k\)}
    \end{cases}
    \notag\\
    &\qquad+\oint_{\partial U}\mathrm d^{d-1}x\,(\dotsb),\notag
\end{align}
where in the last step we have integrated by parts \(k\) times. That is, the moment \(Q^\alpha\) is always conserved precisely when \(l<k\)  (independently of the projector \(P\), in which case \(P\) may be simply the identity) or if the following relation is satisfied when \(l\ge k\):
\begin{equation}\label{eq:expression_that_should_vanish}
    P^\alpha_{j_1\dotso j_la}Y^{i_1\dotso i_ka}_A\int_U\mathrm d^dx\,\delta_{(i_1}^{(j_1}\dotsm\delta^{j_k}_{i_k)}x^{j_{k+1}}\dotsm x^{j_l)}E^A=0.
\end{equation}
The condition for the vanishing of \eqref{eq:expression_that_should_vanish} can be stated in a representation-theoretic language. That is, from the above discussion, the following proposition follows.

\begin{proposition}\label{prop:type_B_criterion}
For a Gauss law \eqref{eq:gauss_law_ansatz},
the conserved moments are either
\begin{itemize}
\item of the form
\begin{equation}\label{conserved_mom_1}
    \int_U\mathrm d^dx\,x^{j_1}\dotso x^{j_l}\rho^a
\end{equation}
for \(l<k\) with \(P=\operatorname{id}\),
\item or of the form
\begin{equation}\label{eq:projection-ansatz}
    P^\alpha_{j_1\dotso j_la}\int_U\mathrm d^dx\,x^{j_1}\dotsm x^{j_l}\rho^a
\end{equation}
for \(l\ge k\),
where
\begin{equation}
    P\colon S\otimes\square^{\odot l} \to \operatorname{coker}(Z)
\end{equation}
is the projection map to the cokernel (i.e.~orthogonal complement to the image) of \(Z\), where \(Z\) is the representation-theoretic branching representing the index contrations that appear in \eqref{eq:expression_that_should_vanish}, that is, the composition
\begin{equation}\label{eq:Z-definition}
    Z\colon R\otimes\square^{\odot(l-k)} \xrightarrow{\operatorname{id}_R\otimes\delta\otimes\operatorname{id}_{\square^{\odot(l-k)}}} R\otimes\square^{\vee k}\otimes\square^{\vee k}\otimes\square^{\odot(l-k)}\xrightarrow{Y\otimes s} S\otimes\square^{\odot l},
\end{equation}
where
\begin{equation}\delta =
\sum_{i=0}^{\lfloor k/2\rfloor}
\operatorname{diag}_{\square^{\vee(k-2i)}}
\in\bigoplus_{i=0}^{\lfloor k/2\rfloor}
\square^{\vee(k-2i)}\otimes\square^{\vee(k-2i)}
\subset\square^{\odot k}\otimes\square^{\odot k}
\end{equation}
is the totally symmetric Kronecker delta \(\delta_{(i_1}^{(j_1}\dotsm\delta_{i_k)}^{j_k)}\), and
\begin{equation}s\colon\square^{\odot k}\otimes\square^{\odot(l-k)}\to\square^{\odot l}\end{equation}
is the total symmetrisation map.
\end{itemize}
\end{proposition}

We may classify Gauss laws according to the following cases.
\begin{itemize}
\item For a \emph{non-fractonic} Gauss law, only the zeroth moment, i.e. the charge \(\int_U\mathrm d^dx\,\rho\) is conserved. This includes the ordinary Gauss law \(\partial_iE^i=\rho\) of Maxwell theory.
\item For a \emph{fractonic} Gauss law, a finite number of higher moments are conserved in addition to the zeroth moment. As discussed in \cite{Pretko:2016kxt,Pretko:2020cko}, this means that single monopole particles are immobile whereas dipoles or other higher-order multipoles are mobile.
\item For a \emph{hyper-fractonic} Gauss law, there is an infinite tower of higher moments that must be conserved.
This renders particle-like dynamics problematic and signals stringy, branelike or other exotic behaviours.
\end{itemize}

In particular, let us consider a simple sufficient (but not necessary) criterion for hyper-fractonicity, whose proof is given in \cref{app:proof}.
\begin{proposition}\label{prop:type_A_criterion}
Consider a Gauss law \eqref{eq:gauss_law_ansatz} that belongs to the special class for which
\begin{equation}
    S \subset \square^{\vee(k-2m)}\otimes R\subset\square^{\odot k}\otimes R.
\end{equation}
This corresponds to Gauss laws containing \(m\) occurrences of the Laplacian \(\Delta\), i.e.
\begin{equation}\label{eq:special_class_gauss_law}
    Y^{i_1\dotso i_ka}_A\partial_{i_1}\dotso\partial_{i_k}E^A=\tilde Y^{i_1\dotso i_{k-2m}a}_A\Delta^m\partial_{i_1}\dotso\partial_{i_{k-2m}}E^A=\rho^a
\end{equation}
where
\begin{equation}
    Y^{i_1\dotso i_ka}
    \sim\tilde Y^{(i_1\dotso i_{k-2m}|a}\delta^{|i_{k-2m+1}i_{k-2m+2}}\dotsm\delta^{i_{k-1}i_k)}.
\end{equation}
Then all moments of the form
\begin{equation}
    \int_U\mathrm d^{d-1}x\,|x|^{2n}x^{\langle l_1}\dotsm x^{j_l\rangle}\rho^a
\end{equation}
are conserved whenever \(n<m\) or \(l+n<k+m\).
\end{proposition}

Therefore, whenever the Gauss law \eqref{eq:gauss_law_ansatz} involves at least one power of \(\Delta\), we always have hyper-fractonic behaviour since there are infinitely many conserved moments of the form \(\int_U\mathrm d^dx\,x^{\langle j_1}\dotsm x^{j_l\rangle}\rho^a\) for arbitrarily large \(l\), according to \eqref{prop:type_A_criterion}. However, as we discuss below, \(p\)-form electrodynamics provides an example of hyper-fractonic behaviour even when the Gauss law does not involve \(\Delta\).

\section{Examples of fractonic behaviour}\label{sec:fract}
We first recapitulate the examples of fractonic behaviour given in \cite{Pretko:2016kxt} in our formalism, enumerating the finite set of conserved moments. This will serve as a testing ground for the results presented in the previous Section.

\subsection{Totally traceless totally symmetric tensor theory}
Let us postulate the Gauss law
\begin{equation}\label{eq:traceless_symmetric_gauss_law}
    \partial_{i_1}\dotsm\partial_{i_k}E^{i_1\dotso i_k} = \rho,
\end{equation}
where \(E^{i_1\dotso i_k}=E^{\langle i_1\dotso i_k\rangle}\) is a totally symmetric totally traceless electric field and \(\rho\) is a scalar charge density. This is a rank-\(k\) generalisation of the rank-two case originally considered in \cite{Pretko:2016kxt} and which admits an action principle \cite{Blasi:2022mbl,Bertolini:2022ijb,Bertolini:2024gzx}.

This is the case where the electric field \(E\) and the charge \(\rho\) follow the \(\operatorname{SO}(d)\) representations
\begin{align}
    R&=\square^{\vee k} &
    S&=\mathbf1,
\end{align}
with the branching from \eqref{eq:Y-definition},
\begin{equation}
    Y\colon \square^{\vee k} \otimes\square^{\dot k}\to \mathbf1.
\end{equation}
The map \(Z\) from \eqref{eq:Z-definition} is
\begin{equation}
    Z\colon\square^{\vee k} \otimes \square^{\odot(l+2n-k)}\xrightarrow{\operatorname{id}\otimes\operatorname{diag}_{\square^{\vee k}}\otimes\operatorname{id}}
    \square^{\vee k} \otimes
    \square^{\vee k}\otimes\square^{\vee k}\otimes
    \square^{\odot(l-k)}
    \to
    \mathbf1\otimes\square^{\vee l},
\end{equation}
so that the cokernel of \(Z\) is given by the cokernel of the total symmetrisation map
\begin{equation}
    \square^{\vee k}\otimes\square^{\odot(l-k)}\to \square^{\odot l}
\end{equation}
between a totally traceless totally symmetric \(k\)-tensor and a totally symmetric \((l-k)\)-tensor into a totally symmetric \(l\)-tensor.
Standard computations show that the cokernel is
\begin{equation}\label{eq:coker_computation}
    \operatorname{coker}(Z)
    =\bigoplus_{n=l-k+1}^{\lfloor l/2\rfloor}\square^{\vee(l-2n)},
\end{equation}
corresponding to conserved moments
\begin{equation}
    Q^{(l-2n,n)i_1\dotso i_l} \coloneqq \int_U\mathrm d^dx\,|x|^{2n}x^{i_1\dotso i_{l-2n}}\rho \qquad(l-k+1\le n\le\lfloor l/2\rfloor).
\end{equation}
In particular, the maximum value of \(l\) for \(\operatorname{coker}(Z)\) \cref{eq:coker_computation} to be nontrivial is \(l=2k-2\), so the number of conserved moments is finite: we have fractonic but not hyper-fractonic behaviour.

In particular, for \(k=1\) (Maxwell theory), we only have the scalar monopole charge \(Q^{(0,0)}\). For \(k=2\) (traceless symmetric 2-tensor), however, we also have conserved vector dipole and scalar quadrupole charges,
\begin{align}
    Q^{(0,0)}&=\int_U\mathrm d^dx\,\rho &
    Q^{(1,0)i}&=\int_U\mathrm d^dx\,x^i\rho &
    Q^{(0,1)}&=\int_U\mathrm d^dx\,|x|^2\rho,
\end{align}
in agreement with the discussion in \cite{Pretko:2016kxt}.

\subsection{Symmetric traceless tensor with a vector charge}
As another example, let us postulate a symmetric traceless electric field
\begin{equation}
    E^{ij}=E^{\langle ij\rangle}
\end{equation}
obeying the Gauss law
\begin{equation}\label{eq:vector-charge-gauss-law}
    \partial_iE^{ij} = \rho^j.
\end{equation}
This is a traceless variant of the example considered in \cite{Pretko:2016kxt}.

According to \cref{prop:type_B_criterion}, the conserved moments are given by the cokernel of the map \(Z\) \eqref{eq:Z-definition}, which in this case is
\begin{equation}
    Z\colon\square^{\vee 2}\otimes\square^{\odot l}\to\square\otimes\square^{\odot l}.
\end{equation}
Computations using e.g.\ Young tableaux or computer algebra systems show that this map is surjective except for \(l\le3\), where the cokernel is
\begin{equation}
    \operatorname{coker}(Z)=\begin{cases}
        \square & \text{if \(l=2\)}\\
        \mathbf1\oplus\square^{\wedge2}&\text{if \(l=1\)} \\
        \square & \text{if \(l=0\)}.
    \end{cases}
\end{equation}
These correspond to the four conserved moments
\begin{align}
    Q^{(0)}&\coloneqq\int_U\mathrm d^dx\,\rho&
    Q^{(1)}&\coloneqq\int_U\mathrm d^dx\,x\cdot\rho\\
    Q^{(1')ij}&\coloneqq\int_U\mathrm d^dx\,x^{[i}\rho^{j]}&
    Q^{(2)i}&\coloneqq
    \int_U\mathrm d^dx\,\left[|x|^2\rho^i-2(\rho\cdot x)x^i\right].
\end{align}
Since we only have a finite number of conserved moments, the system is fractonic but not hyper-fractonic.

\section{Examples of hyper-fractonic behaviour}\label{sec:hyper_fracton}
In this section, we will analyse examples of new fractonic behaviour that arise from the analysis of Section~\ref{sec:general_conserved_moments}.
\subsection{Polyharmonic equation}\label{ssec:polyharmonic}
As a first simple (but perhaps physically unrealistic) example of hyper-fractonic behaviour, we can consider the Gauss law given by the inhomogeneous polyharmonic equation
\begin{equation}\label{eq:gauss_law_scalar}
    \Delta^mE=\rho,
\end{equation}
i.e.\ when in \eqref{eq:gauss_law_ansatz} when we have
\begin{equation}
    Y\colon \square^{\odot(2m)}\otimes\mathbf1\to\mathbf1,
\end{equation}
and where both the electric field \(E\) and the charge \(\rho\) are scalars.

Let us consider the conserved moments of \eqref{eq:gauss_law_scalar} according to \cref{prop:type_B_criterion}. This is the case where \(R=S=\mathbf1\) is the singlet representation. The branching \(Y\colon\mathbf1\otimes\mathbf1\to\mathbf1\) from \eqref{eq:Y-definition} is trivial. The map \(Z\) \eqref{eq:Z-definition} is given by
\begin{equation}
    Z\colon\square^{\odot(l-2m)}\to\square^{\odot l}.
\end{equation}
Since
\begin{align}
    \square^{\odot l} &= \square^{\vee l}\oplus\square^{\vee(l-2)}\oplus\dotsb \\
    \square^{\odot(l-2m)} &= \square^{\vee(l-2m)}\oplus\square^{\vee(l-2m-2)}\oplus\dotsb,
\end{align}
we see that
\begin{equation}
    \operatorname{coker}(Z) = \square^{\vee l}\oplus\square^{\vee(l-2)}\oplus\dotsb\oplus\square^{\vee(l-2m+2)}.
\end{equation}
This corresponds to the conserved moments
\begin{equation}\label{eq:polyharmonic_conserved_moments}
    Q^{(l,n)i_1\dotso i_l}=\int_U\mathrm d^dx\,|x|^nx^{\langle i_1}\dotsm x^{i_l\rangle}\rho\qquad(l\in\mathbb Z_{\ge0},\;n\in\{0,\dotsc, m-1\}).
\end{equation}
As \(l\) is unbounded, we always have hyper-fractonic behaviour.

For \(m=1\), for example, \eqref{eq:gauss_law_scalar} is simply the Poisson (inhomogeneous Laplace) equation, which was discussed in \cite{Hart:2021gre}, where an infinite number of conservation laws emerge from the linearised hydrodynamic theory of isotropic dipole-conserving fluids. The conserved moments correspond to harmonic polynomials \(x^{i_1\dotso i_k}\):
\begin{align}
    Q^{i_1\dotso i_k} &= \int_U\mathrm d^dx\,x^{i_1\dotso i_k}\rho.
\end{align}

\subsection{\(p\)-form electrodynamics}
Consider a theory of \(p\)-form electrodynamics with action
\begin{equation}
    S = \int\mathrm d^{d+1}x\,\left(-\frac1{2(p+1)!}F_{\mu_0\mu_1\dotso\mu_p}F^{\mu^0\mu^1\dotso\mu^p}
    + \frac1{p!}A_{\mu_1\dotso\mu_p}J^{\mu_1\dotso\mu_p}\right),
\end{equation}
where \(A_{\mu_1\dotso\mu_p}=A_{[\mu_1\dotso\mu_p]}\) is a \(p\)-form gauge field, \(J^{\mu_1\dotso\mu_p}=J^{[\mu_1\dotso\mu_p]}\) is a \(p\)-form current that couples to it, and the field strength is
\begin{equation}
    F_{\mu\nu_1\dotso\nu_p} = (p+1)\partial_{[\mu}A_{\nu_1\dotso\nu_p]}.
\end{equation}
The equations of motion are then
\begin{equation}
    \partial^\mu F_{\mu\nu_1\dotso\nu_p} + J_{\nu_1\dotso\nu_p} = 0.
\end{equation}
In particular, defining the electric field and charge as
\begin{align}
    E_{i_1\dotso i_p} &\coloneqq F_{0i_1\dotso i_p} &
    \rho_{i_1\dotso i_{p-1}} &\coloneqq J_{0i_1\dotso i_{p-1}},
\end{align}
we have the Gauss law
\begin{equation}
    \partial^jE_{ji_1... i_{p-1}}=\rho_{i_1\dotso i_{p-1}}.
\end{equation}
The case \(p=1\) is that of ordinary Maxwell theory.
However, for \(p>1\), according to \cref{prop:type_B_criterion} these systems admit infinitely many conserved moments and are therefore hyper-fractonic: an infinite tower of conserved moments is given by
\begin{equation}
    Q^\alpha=P^\alpha_{i_1\dotso i_{p-1}j_1\dotso j_l}\int_U\mathrm d^dx\,\rho^{i_1\dotso i_{p-1}}x^{\langle j_1}\dotsm x^{j_l\rangle},
\end{equation}
where \(P\) is the projection to the \(\operatorname{SO}(d)\)-representation \(\square^{\wedge(p-1)}\vee\square^{\vee l}\), that is, the highest-weight component of \(\square^{\wedge(p-1)}\otimes\square^{\vee l}\).

At first sight, the appearance of hyper-fractonic behaviour for \(p\)-form electrodynamics, with its infinite number of constraints, would seem to preclude dynamical charges. However, \(p\)-form potentials naturally couple to \((p-1)\)-branes, which may be dynamical.
That is, while a finite number of point charges cannot move due to the hyper-fractonic constraints, a continuum of point charges that arrange themselves into a string can. That is, in this case the hyper-fractonic constraints signal the appearance of extended objects.

Note that \(p\)-form electrodynamics meets the assumptions of the Coleman--Mandula theorem \cite{Coleman:1967ad,Mandula:2015}, which requires that all conserved moments vanish since their generators do not commute with Poincaré symmetry. For example, for \(p=2\), if one has a closed string (with current density \(J_{\mu\nu}\) and hence charge density \(\rho_i\sim J_{0i}\)) in some compact spatial region \(U\), one has
\begin{equation}\label{eq:closed_string_discussion}
    \int_U\mathrm d^dx\,J_{0i} = 0.
\end{equation}
This differs from the usual discussion of higher-form symmetries \cite{Gomes:2023ahz,Cordova:2022ruw,Schafer-Nameki:2023jdn,Luo:2023ive,Bhardwaj:2023kri,Brennan:2023mmt}. Here we are integrating against \(d\)-dimensional spatial volume (in \(d+1\)-dimensional spacetime), whereas for higher-form symmetries, we would be integrating against a (one-dimensional) curve in space (or a two-dimensional worldsheet in spacetime).

For concreteness, let us consider the \(p=2\) case, such as the Kalb--Ramond field in Type~2 supergravity \cite{Freedman:2012zz}, where \(R=\square^{\wedge2}\) and \(S=\square\). For an \(l\)th-order moment (\(l\ge0\)), the map \(Z\) is
\begin{equation}
    Z\colon\square^{\wedge2}\otimes\square^{l-1}\to\square\otimes\square^{\odot l}.
\end{equation}
For \(l=0\), we have the zeroth moment (i.e.~charge)
\begin{equation}
    Q^{(0)i}=\int_U\mathrm d^dx\,\rho^i.
\end{equation}
For \(l=1\) (i.e.\ a dipole moment), the map \(Z\) is
\begin{equation}
    Z\colon\square^{\wedge2} \to \square\otimes\square=\square^{\wedge2}\oplus\square^{\odot2},
\end{equation}
whose cokernel is
\begin{equation}
    \operatorname{coker}(Z)=\square^{\odot2}=\square^{\vee2}\oplus\mathbf1.
\end{equation}
This corresponds to the dipole moment
\begin{align}
    Q^{(1)ij}&\coloneqq\int_U\mathrm d^dx\,x^{(i}\rho^{j)}.
\end{align}
For \(l=2\) (i.e.\ quadrupole moments), the map \(Z\) is
\begin{equation}
\begin{aligned}
    Z\colon&\square^{\wedge2}\otimes\square=\square\oplus \square^{\wedge3}\oplus(\square^{\wedge2}\vee\square)\\
    &\to \square\otimes(\square^{\vee2}\oplus\mathbf1)=\square\oplus\square\oplus \square^{\vee3}\oplus(\square^{\wedge2}\vee\square).
\end{aligned}
\end{equation}
We see that the cokernel is \(\square\oplus\square^{\vee3}\), which corresponds to the conserved vector and symmetric rank-3 quadrupole moments
\begin{align}
    Q^{(2)i} &= \int_U\mathrm d^dx\,\left(x^i(\rho\cdot x)+\frac12\rho^i|x|^2\right) \\
    Q^{(2)ijk} &= \int_U\mathrm d^dx\,\rho^{\langle i}x^jx^{k\rangle}.
\end{align}
For \(l=3\) (i.e.\ octupole moments),
\begin{equation}
\begin{aligned}
    Z\colon&\square^{\wedge2}\otimes(\square^{\vee2}\oplus\mathbf1) =\square^{\wedge2}\oplus\square^{\wedge2}\oplus(\square^{\wedge3}\vee\square)\oplus\square^{\vee2}\oplus(\square^{\vee2}\vee\square^{\wedge2})\\
    &\to \square\otimes(\square^{\vee3}\oplus\square)
    =\mathbf1\oplus\square^{\wedge2}\oplus\square^{\vee2}\oplus\square^{\vee2}\oplus(\square^{\vee2}\vee\square^{\wedge2})\oplus\square^{\vee4}.
\end{aligned}
\end{equation}
Hence the cokernel of \(Z\) is \(\mathbf1\oplus\square^{\vee2}\oplus\square^{\vee4}\), whose corresponding conserved octupole moments are
\begin{align}
    Q^{(3)}&=\int_U\mathrm d^dx\,(\rho\cdot x)|x|^2\\
    Q^{(3)ij}&=\int_U\mathrm d^dx\,\left((\rho\cdot x)x^{\langle i}x^{j\rangle}+\rho^{\langle i}x^{j\rangle}|x|^2\right)\\
    Q^{(3)ijkl}&=\int_U\mathrm d^dx\,\rho^{\langle i}x^jx^kx^{l\rangle}.
\end{align}
In general, the totally symmetric totally traceless moments
\begin{equation}
    Q^{(l)i_0\dotso i_l}\coloneqq\int_U\mathrm d^dx\,\rho^{\langle i_0}x^{i_1}\dotsm x^{i_l\rangle}
    =\int_U\mathrm d^dx\,P^{i_0\dotso i_l}_{j_0\dotso j_l}\rho^{j_0}x^{j_1}\dotsm x^{j_l}
\end{equation}
are always conserved for any \(l\) according to \cref{prop:type_B_criterion}, where the projection
\begin{equation}
    P\colon \square^{\odot(l+1)}\to \square^{\vee(l+1)}
\end{equation}
is projection to the totally traceless totally symmetric part, but there are many other conserved moments in addition.

\subsection{Exhaustive enumeration in low rank}\label{ssec:low_rank_scan}

Using representation theory, we may easily enumerate Gauss laws of lower ranks.
\begin{table}[H]
\begin{center}
\begin{tabular}{cccc}\toprule
Gauss law & Behaviour \\\midrule
\(\partial^i E=\rho^i\) & hyper-fractonic\\
\(\Delta E = \rho\) & hyper-fractonic\\
\(\partial^{\langle i}\partial^{j\rangle} E = \rho^{ij} \) & hyper-fractonic\\
\(\partial_iE^i = \rho\) & non-fractonic \\
\(\partial^{[i}E^{j]} = \rho^{ij}\) & hyper-fractonic\\
\(\partial^{\langle i}E^{j\rangle} = \rho^{ij}\) & hyper-fractonic \\
\(\Delta E^i = \rho^i\) & hyper-fractonic \\
\(\partial^{\langle i}\partial^{j\rangle}E_j = \rho^i \) & hyper-fractonic \\
\(\partial^{\langle i}\partial^jE^{k\rangle}=\rho^{ijk}\) & hyper-fractonic \\
\((Y_{\square^{\wedge2}\vee\square})^{ijk}_{i'j'k}\partial^{\langle i'}\partial^{j'\rangle} E^{k'}=\rho^{ijk}\) & hyper-fractonic \\
\bottomrule
\end{tabular}
\end{center}

\caption{\footnotesize{Gauss laws involving up to two derivatives and field strengths with up to one index. In the above,
\(Y_{\square^{\wedge2}\vee\square}\colon\square^{\vee2}\otimes\square\to\square^{\wedge2}\vee\square\)
is the projector to an irreducible subrepresentation.}}\label{table:low_rank}
\end{table} \Cref{table:low_rank} enumerates the ten possible Gauss laws involving at most two derivatives and a field strength \(E\) carrying at most one Lorentz index.

Note that we only enumerate Gauss laws involving electric fields, charges and differential operators that are irreducible representations of \(\operatorname{SO}(d)\). For example, the Gauss law
\begin{equation}
    \partial_i\partial_jE^j = \rho_i
\end{equation}
is not included, since it can be seen as the composition of the following two Gauss laws
\begin{align}
    \Delta E^i &= \rho^i &
    \partial^{\langle i}\partial^{j\rangle}E_j&= \rho^i,
\end{align}
which are included in \cref{table:low_rank}.

\section{Discussion}\label{sec:final_remarks}
In this paper, we have shown that a generalised Gauss law \eqref{eq:gauss_law_ansatz} in arbitrary dimensions allows for two kinds of possible conserved moments, \eqref{conserved_mom_1} and/or \eqref{eq:projection-ansatz}, from which we can observe three behaviours:
	\begin{enumerate}
	\item non-fractonic : only the charge is conserved and it has complete mobility~;
	\item fractonic : a finite number of higher moments is conserved, which present restricted mobility (complete or partial)~;
	\item hyper-fractonic : an infinite tower of conserved moments, which may signal stringy/brane-like behaviour.
	\end{enumerate}
The most surprising result that emerges from our analysis is the last case: a new general behaviour presenting an infinite tower of conserved charges. As unphysical and unrealistic as it may seem, a particular case of that was observed in \cite{Hart:2021gre} in the context of hydrodynamics and fractons. This provides a confirmation of our procedure, which then brought us to identify this feature also in higher-form gauge theories, which thus seem to exhibit a new (hyper-)fractonic behaviour. In addition, the analysis presented herein may then provide a guiding light to construct new (hyper-)fractonic theories. In fact, another way to obtain multipole conservation is through continuity equations of matter sources, which reflect the Gauss laws of the specific theory. For instance, in the scalar-charge theory, it is
	\begin{equation}\label{cont}
	    \partial_0\rho+\partial_i\partial_jJ^{ij}=0,
	\end{equation}
which, when integrated over a volume, implies that the total dipole must be conserved in time (up to boundary terms) \cite{Pretko:2016kxt,Pretko:2016lgv}. \Cref{cont} may then be seen as a Ward identity on the sources, from which it is thus possible to derive the symmetry transformation of the associated gauge field, from which finally the invariant action and the full theory can be reconstructed.

\appendix
\section{Proof of \cref{prop:type_A_criterion}}\label{app:proof}
Consider the Gauss law of the form \eqref{eq:special_class_gauss_law}, viz.
\begin{equation}
    \tilde Y^{ai_1\dotso i_k}_A\Delta^m\partial_{i_1}\dotso\partial_{i_k}E^A = \rho^a.
\end{equation}
Then, integrating by parts to obtain a conservation law, the following expression appears:
\begin{equation}
    \Delta^m\partial_{\langle i_1}\dotsm\partial_{i_k\rangle}(x^{j_1}\dotsm x^{j_{1+2n}}).
\end{equation}
Decomposing the string of \(x\)s into irreducible representations of \(\operatorname{SO}(d)\), it suffices to show that
\begin{equation}\label{eq:kronecker_irrep}
    \Delta^m\partial_{\langle i_1}\dotsm\partial_{i_k\rangle}(|x|^{2n}x^{\langle j_1}\dotsm x^{j_l\rangle}).
\end{equation}
vanishes precisely when either \(n<m\) or \(l+n<k+m\), since this is the expression that remains when one integrates by parts the conserved charge.

The expression \eqref{eq:kronecker_irrep}
may be represented as a suitable projection
\begin{equation}\label{eq:proj1}
    \Delta^m\partial_{\langle i_1}\dotsm\partial_{i_k\rangle}(|x|^{2n}x^{\langle j_1}\dotsm x^{j_l\rangle})
    =
    P^{i'_1\dotso i'_{k+2m}}_{i_1\dotso i_k}
    {\tilde P}_{j'_1\dotso j'_{l+2m}}^{j_1\dotso j_l}\partial_{i'_1}\dotso\partial_{i'_{k+2m}}(x^{j'_1}\dotsm x^{j'_{l+2n}})
\end{equation}
where \(P\) and \(\tilde P\) are suitable projectors to subrepresentations:
\begin{align}
    P\colon \square^{\odot(k+2m)}&\to\square^{\vee k} &
    \tilde P\colon \square^{\odot(l+2n)}&\to\square^{\vee l}.
\end{align}
It is easy to see that
\begin{equation}\label{eq:kronecker_full}
    \partial_{i_1}\dotsm\partial_{i_{k+2m}}(x^{j_1}\dotsm x^{j_{l+2n}})
    = \begin{cases}
        \binom lk\delta_{(i_1}^{(j_1}\dotsm\delta_{i_{k+2m})}^{j_{k+2m}}x_{\vphantom{i_k}}^{j_{k+2m+1}\dotso j_{l+2n})} & \text{if \(l+2n\ge k+2m\)} \\
        0 &\text{if \(l+2n < k+2m\)}.
    \end{cases}
\end{equation}
That is, the expression \(\partial_{i'_1}\dotso\partial_{i'_{k+2m}}(x^{j'_1}\dotsm x^{j'_{l+2n}})\) is either zero (if \(k+2m>l+2n\)) or a product of \(l+2n-k-2m\) occurrences of \(x\) followed by some Kronecker deltas.

The string of \(x\)s in the right-hand side of \eqref{eq:kronecker_full} transforms as the representation \(\square^{\odot(l+2n-k-2m)}\). The Kronecker deltas (after separating out various traces) are the diagonal map \(\operatorname{diag}_{\square^{\wedge k}}\) on the representation \(\square^{\wedge k}\), and \(\operatorname{diag}_{\square^{\wedge k}}\) transforms as \(\square^{\vee k}\otimes\square^{\vee k}\).

So \eqref{eq:proj1} may be evaluated by (a) starting with a string of \(x\)s, (b) contracting with Kronecker deltas, and (c) projecting out some traces.
That is, the projection involved here is
\begin{equation}
\begin{aligned}
    \square^{\odot(l+2n-k-2m)}
    &\xrightarrow{\operatorname{diag}_{\square^{\vee k}}\otimes\operatorname{id}_{\square^{\odot(l+2n-k-2m)}}}\square^{\vee k}\otimes\square^{\vee k}\otimes\square^{\odot(l+2n-k-2m)}\\
    &\xrightarrow{\operatorname{id}_{\square^{\vee k}}\otimes s}\square^{\vee k}\otimes\square^{\vee l}
\end{aligned}
\end{equation}
where
\begin{equation}\label{eq:symmetrisation_map}
    s\colon\square^{\vee k}\otimes\square^{\odot(l+2n-k-2m)}\to\square^{\vee l}.
\end{equation}
So, the criterion for \eqref{eq:kronecker_irrep} vanishing is whether
\begin{equation}
    \square^{\vee k}\otimes\square^{\odot(l+2n-k-2m)}
    \overset?{\not\supset}\square^{\vee l}.
\end{equation}
This is equivalent to
\begin{multline}
    l\not\in \{k-(l+2n-k-2m),k-(l+2n-k-2m)+2,k-(l+2n-k-2m)+4,\\\dotsc,k+(l+2n-k-2m)\},
\end{multline}
which simplifies to
\begin{equation}\label{eq:kronecker_irrep_criterion}
    m>n\text{ or }k+m >l+n,
\end{equation}
as desired.

\section*{Acknowledgements}
The authors thank Leron Borsten\orcidlink{0000-0001-9008-7725}, Rahul Mahajan Nandkishore\orcidlink{0000-0001-5703-6758}, and Gian\-domenico Palumbo\orcidlink{0000-0003-1303-1247} for helpful comments.

\bibliographystyle{unsrturl}
\bibliography{biblio,biblio2}

\end{document}